\documentclass[manuscript=article,journal=jpclcd]{achemso}

\usepackage{amsmath}
\usepackage{graphicx}
\usepackage{amssymb}
\usepackage[usenames]{color}
\usepackage{subfigure}
\usepackage{dcolumn}
\usepackage{bm}
\usepackage{ulem}
\usepackage{color}
\bibstyle{achemso}
\setkeys{acs}{usetitle = true}

\newcommand{\ek}{ \langle E_K\rangle}
\newcommand{\ndp}{$n$($p$) }

\makeatletter

\title{Direct measurements of quantum kinetic energy tensor in stable and metastable water near the triple point: an experimental benchmark}

\author{Carla Andreani}
\affiliation {Universit\`{a} degli Studi di Roma "Tor Vergata", Dipartimento di Fisica e Centro NAST, Via della Ricerca Scientifica 1, 00133 Roma, I}
\affiliation {Consiglio Nazionale delle Ricerche, CNR-IPCF, Sezione di Messina, I}
\author{Giovanni Romanelli}
\affiliation {ISIS Neutron Source, Science Technology Facility Council, Chilton, Oxfordshire, OX11 0QX, UK}
\author{Roberto Senesi}
\email{roberto.senesi@uniroma2.it}
\affiliation {Universit\`{a} degli Studi di Roma "Tor Vergata", Dipartimento di Fisica e Centro NAST, Via della Ricerca Scientifica 1, 00133 Roma, I}
\affiliation {Consiglio Nazionale delle Ricerche, CNR-IPCF, Sezione di Messina, I}

\begin{document}
\singlespacing
\begin{abstract} 
This study presents the first direct and quantitative measurements of the nuclear momentum distribution anisotropy and the quantum kinetic energy tensor in stable and metastable (supercooled) water near its triple point using Deep Inelastic Neutron Scattering (DINS). From the experimental spectra accurate lineshapes of the hydrogen momentum distributions are derived using an anisotropic Gaussian and a model independent framework. The experimental results, benchmarked with those obtained for the solid phase, provide the state of the art directional values of the hydrogen mean kinetic energy in metastable water. The determinations of the direction kinetic energies in the supercooled phase, benchmarked with ice at the same temperature, provide accurate and quantitative measurements of these dynamical observables in metastable and stable phases, {i.e.} key insight in the physical mechanisms of the hydrogen quantum state in both disordered and polycrystalline systems. The remarkable findings of this study establish novel insight to further expand the capacity and accuracy of DINS investigations of the nuclear quantum effects in water and represent reference experimental values for theoretical investigations.

 
\end{abstract}

\maketitle

{\bf TOC Graphic:}\\
\includegraphics[width=8cm]{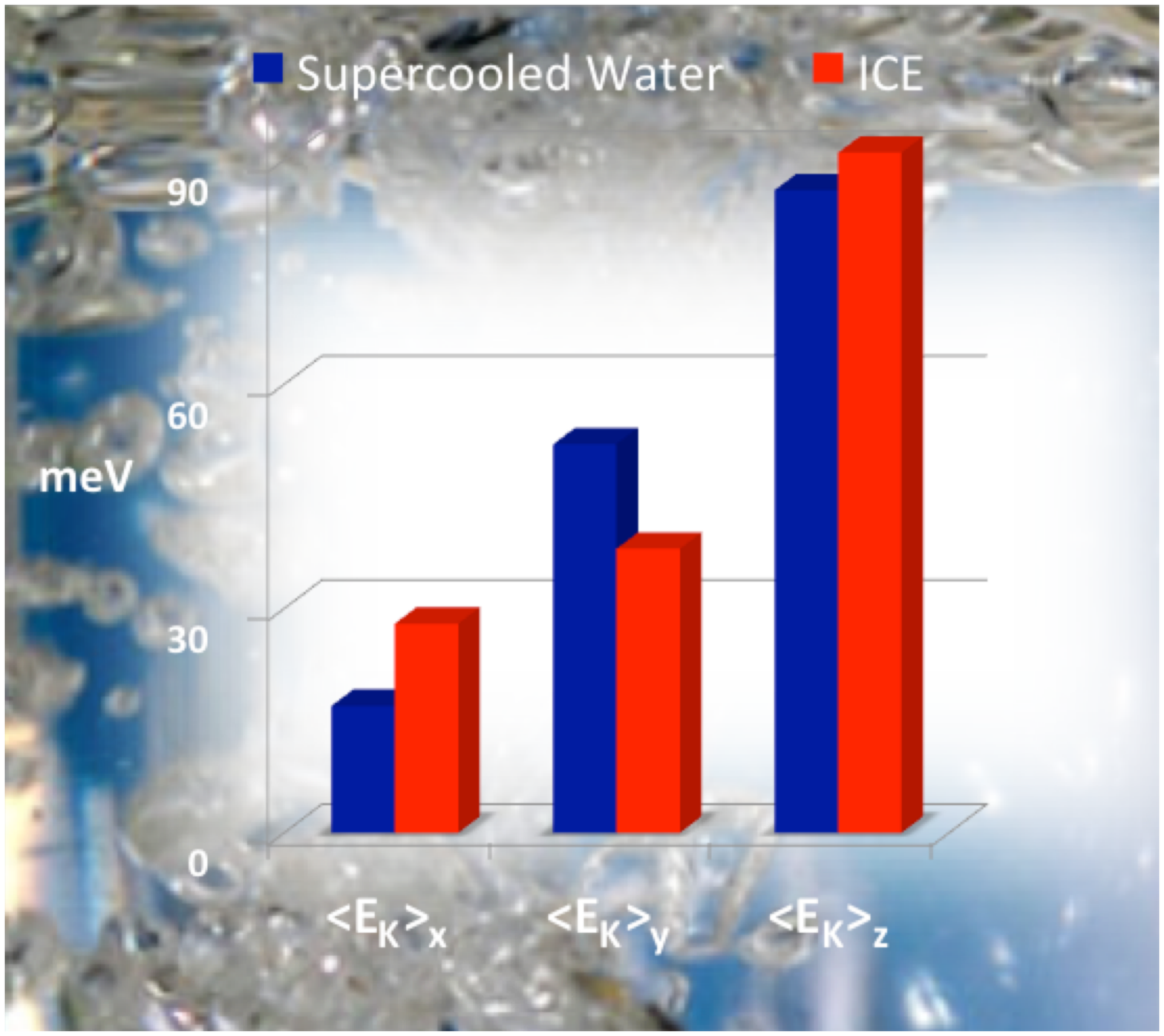}
\\
{\em keywords:} deep inelastic neutron scattering; ab initio path integral molecular dynamics; particle momentum distribution
\\


A large number of experimental and theoretical dynamical studies of liquid water near the triple point are available in literature\cite{2004Reiter_bjp,2006burnham,morrone_2007,2008Pantalei_prl,Burnham_2008,giuliani2011isotope,ramirez2011kinetic,Bruni_2012,2014moreh} nevertheless an full and accurate characterization of hydrogen dynamics is still lacking. The latter is of vital importance for clarifying thermodynamic properties and the key to expand our understanding of some of the mysterious characteristics of water, supercooled water (SW) and glassy water, the latter being its viscous counterparts, known as amorphous ice\cite{Parmentier}. 

Nuclear Quantum Effects (NQEs) play an important role in water, ice and hydrogen-bonded systems and directly influence their microscopic structure and dynamical properties. In most of these cases, the hydrogen atoms are localized in potential well with pronounced zero point motion. The equilibrium hydrogen dynamics is reflected in the quantum momentum distribution, $n(p)$, a quantity which provides complementary information to what is garnered from diffraction techniques. Due to NQEs \ndp markedly differs from the classical Maxwell-Boltzmann distribution, and is determined almost entirely by the quantum mechanics of the vibrational ground state properties \cite{2005andreani_adv,1966hohemberg,1957goldanskii,1964ivanov,Reiter_2002,2010reiterPRL,2011farad}. This makes \ndp a highly sensitive probe of the local environment, fingerprinting any changes occurring both in the structure of the hydrogen-bonded network as well as in the local symmetry. Thus \ndp together with the mean kinetic energy, $\ek$, provide unique key insights into the hydrogen local environment to rationalize the puzzling feature of liquid water near the triple point. DINS is the unique experimental technique that directly access the $n(p)$\cite{1986_Gunn}. The basic principles of data interpretation of the DINS technique are based on the validity of the Impulse Approximation (IA)\cite{1975west} which is exact in the limit of infinite momentum transfer, $\hbar q$\cite{Reiter_1985, Watson_1996}. Within the IA, the inelastic neutron scattering cross-section directly probes the \ndp of each nucleus in the target system\cite{2005andreani_adv}. Many DINS experiments have uncovered the hydrogen \ndp in a large variety of  water systems showing how its line shape fingerprints the change in the hydrogen network. Recent examples are DINS in water and ice reflecting the breaking and distortion of the hydrogen bonds that occurs upon melting \cite{2013andreani} or the competition between intra and inter molecular NQEs unraveled by measuring the anisotropy of the quantum kinetic energy tensor of D and O in D$_2$O\cite{2013romanellijpcl,Parmentier}.  
Recent Inelastic Neutron Scattering (INS) studies of ice, water, SW and amorphous ices (Low Density, High Density and Very High Density) measurements show that, in an harmonic framework, almost identical value of $\ek_z$, the OH stretching component of the $\ek$, are obtained: 98 meV, 100 meV, 100 meV and 99-101 meV, respectively\cite{2013senesi_OH_str_jcp, Parmentier}. This is an indication that the NQEs on the OH stretching frequency, $\omega_z$, is weakly dependent in the temperature range explored. On the contrary DINS measurements go beyond the harmonic framework and provide full directional components of the mean kinetic energy tensor\cite{Parmentier}.

From the theoretical point of view empirical flexible and polarizable models\cite{2006burnham,Burnham_2008} are not able to fully capture the small difference between the $n(p)$ distributions of water near its triple point one observes in DINS and INS experiment. First principles molecular dynamics such as open Path Integral Car-Parrinello Molecular Dynamics (PICPMD), unlike less transferable models, is a promising path for the exploration of these detailed features in \ndp  and $\ek$\cite{morrone_2007}. In particular in a DINS experiment on ice the quantum $\ek$ has been measured on ice at 271 K, and the observable has been used as a quantitative benchmark for electronic density functionals employed using a PICPMD in the description of hydrogen bonded systems in {\it ab initio} numerical simulations\cite{2011Flammini}. Beyond ice, path integral simulation studies are employed to study diluted water phase such as supercritical water\cite{2008Pantalei_prl}. In these cases experimental and theoretical predictions of $\ek$ are in satisfactory quantitative agreement. On the other hand discrepancies still exists between experiments and theory for liquid water at room temperatures and below. Particularly significant is the excess of $\ek$ across the density maximum at 277 K and across the supercooled phase at 272 K observed in DINS experiments\cite{2008_pietropaolo_prl,BZ_2009,2009flammini}. Similar excess of $\ek$ is also found in supercooled heavy water\cite{giuliani2011isotope} and highly pressurized SW\cite{Bruni_2012}. The interpretation of these experiments is still matter of debate. Indeed in the case of SW at T=271 K an excess of $\ek$, about 58$\%$ (8 kJ/mol) higher with respect to the room-temperature result\cite{2008_pietropaolo_prl,BZ_2009,2009flammini}, seems to be incompatible with the experimental molecular structure of water. Yet the proposed explanations are not supported by any computer simulation calculation or theoretical model. Very recent path integral molecular dynamics provide accurate results for nuclear momentum distribution and \cite{ceriotti1} of water near the triple point. Thus accurate DINS experimental determination are advisable in order to allow an quantitative cross comparison between DINS experimental and simulation determinations of the \ndp. 

\begin{figure}
\centering\includegraphics[width=14cm]{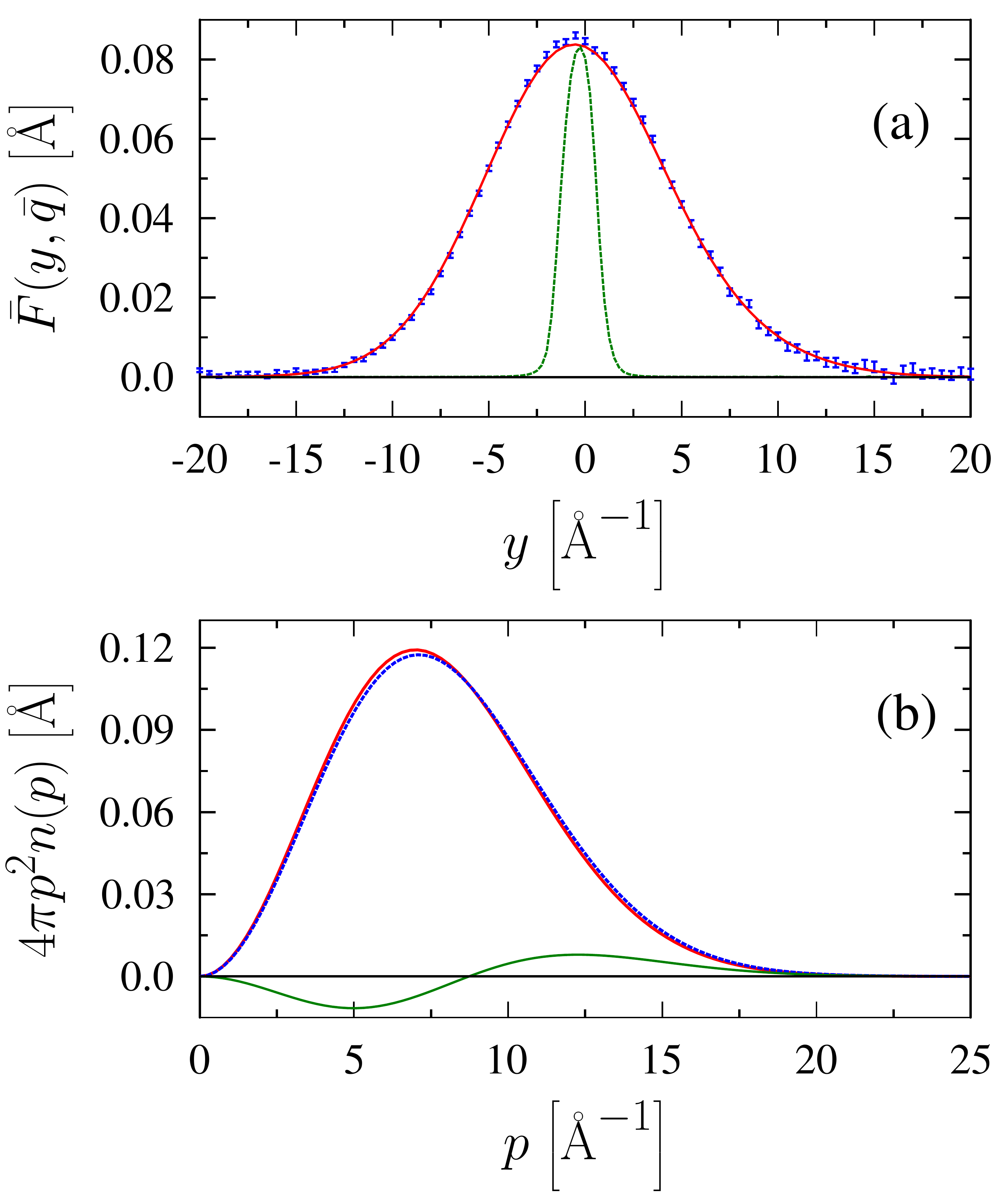}
\caption{(colour online) (a) Angle averaged hydrogen NCP $\bar F(y,q)$ for SW at $T$ = 271 K {\color{blue}blue error bars}. The best fit of this function using M2 is plotted as {\color{red}red line}. The experimental resolution $R(y,q)$ is plotted as  {\color{green}green line}. (b) Radial momentum distributions $4\pi p^2n(p)$ for SW ({\color{blue}blue line}) and ice ({\color{red}red line}) at $T$ = 271 K. The difference between SW and ice line-shapes (magnified by a factor of 10) is plotted as a {\color{green}green line}.}
\label{cy_data}
\end{figure}

In this paper we present a DINS study of water near the triple point to determine accurate measurements of hydrogen \ndp and hydrogen directional mean kinetic energy components, $\ek_\alpha$. The single particle dynamics in SW and ice at T=271 K and water at 300 K, is investigated employing new experimental and cooling set up. The measurement on ice is specifically recorded at the same temperature used in Ref.\cite{2011Flammini} in order to allow a quantitative benchmark with measurement on SW. The DINS measurements are carried out on the VESUVIO beamline at the ISIS pulsed neutron and muon source (Rutherford Appleton Laboratory, Chilton, Didcot, UK)\cite{Vesuvio,2011APRSphysrep,2012Mayers,2014seel}. 
In the DINS experiment we obtain, from each {\it l-th} detector, a Neutron Compton Profile (NCP) for the hydrogen nuclei, $F_l (y,q)$. These functions represent the hydrogen longitudinal momentum distribution. Full details on DINS formalism, detailed description of operation of VESUVIO instrument, sample preparation and experimental set up, procedure of cooling and phase monitoring, measurements and data analysis are reported in the Supporting Information.
The experimental angular average of the $F_l (y,q)$ functions, namely $\bar F(y,q)$, for SW at $T$ = 271 K is plotted in the top panel of Figure \ref{cy_data}, together with the angle averaged experimental resolution function, $\bar R(y,q)$ (for full detail see SM). Figures of similar statistical accuracy have been obtained for the other DINS data sets of ice and water samples.

Primary goal of this study is to derive the $n(p)$ lineshapes, $\ek$ and its directional components, $\ek_\alpha$, from the set of $F$($y,q$) spectra. For each sample a simultaneous fit of the individual $F_l(y,q)$ spectra is accomplished using two parametric models for the $n(p)$: (a) a model-independent lineshape, hereafter named Model 1 (M1) and (b) a three dimensional anisotropic Gaussian lineshape derived from a quasi harmonic model, hereafter named Model 2 (M2). The latter is recently employed to reveal the local environment of hydrogen in polycrystalline ice \cite{2011Flammini,2013andreani}, amorphous ice\cite{Parmentier} and heavy water \cite{2013romanellijpcl}. Then experimental $n(p)$ in M1 is given by the Gauss-Laguerre expansion\cite{Reiter_2002,Reiter_1985,2008_pietropaolo_prl}
\begin{equation}
n_{M1}(p)=\frac{\exp\left(-\frac{p^2}{2\sigma^2}\right)}{(\sqrt{2\pi}\sigma)^3}\sum_n c_n (-1)^n L_n^{\frac12}\left(\frac{p^2}{2\sigma^2}\right),
\label{defnp} 
\end{equation}
where $L^{\frac{1}{2}}_{n}$ are the generalized Laguerre polynomials, and $c_n$ the expansion coefficients from which, together with the standard deviation, $\sigma$, one can derive the momentum distribution lineshape.  

\begin{table}
\centering
\begin{tabular}{lccccc}
\hline\hline
			&			 &		SW	&	Ice 		& 	Ice\cite{2011Flammini} 	\\
		T		&	[K]		 &		 271  	&	271  		& 	 271  	\\

\hline
$\bf M1$		&			 &			&		&& 	\\
$\sigma	$		&[\AA$^{-1}$]&		5.01$\pm$0.02 	&	5.03$\pm$0.03		&5.01$\pm$0.03	\\
$c_4$			&			 &		0.11$\pm$0.01	&	0.11$\pm$0.02		&0.10$\pm$0.01	\\
$\ek$			&[meV]		 &		{\color{blue}156.0$\pm$2.0	}&	{\color{red}157.0$\pm$2.0}		&156.0$\pm$2.0 &			\\
\hline
$\bf M2$				&			 &			&		&& 	\\
$\sigma_x$		&[\AA$^{-1}$]&		2.9$\pm$0.5	&	3.7$\pm$0.1 		&3.7$\pm$0.3&		\\
$\sigma_y$		&[\AA$^{-1}$]&		5.0$\pm$0.5	&	4.3$\pm$0.3 		&4.3$\pm$0.4&		\\
$\sigma_z$		&[\AA$^{-1}$]&		6.5$\pm$0.2	&	6.6$\pm$0.2 		&6.5$\pm$0.4& 	\\
$\ek_x$			&[meV]		 &		17$\pm$5	&	28$\pm$2	 		&29$\pm$4&	 	\\
$\ek_y$			&[meV]		 &		52$\pm$10 	&	38$\pm$5 			&38$\pm$9
&	 	\\
$\ek_z$			&[meV]		 &		86$\pm$5		&	91$\pm$5 			&87$\pm$9&  	\\
$\ek$			&[meV]		 &		{\color{blue}156.0$\pm$2.0	}&	{\color{red}157.0$\pm$2.0}		&154.0$\pm$2.0&		\\
\hline\hline
\end{tabular}
\caption{(colour online) The $\ek$ and individual $\ek_\alpha$ values, from present DINS measurements in bulk SW and ice at $T$ = 271 K are shown. These are obtained using M2, while $\ek$, $c_4$ and $\sigma$ values using M1. The latter is found to be equal to $\bar\sigma=\sqrt{\sum\sigma_\alpha^2/3}$ where the $\sigma_\alpha$ are from M2.}
\label{fit_parameters}
\end{table}

The experimental $n(p)$ lineshapes in M2 is determined by modeling the momentum distributions as spherical averages of multivariate
Gaussians according to\cite{2011Flammini}
\begin{equation}  
4\pi p^2  n_{M2}(p)=\Big\langle \frac{\delta(p-|\mathbf{p}|)}{\sqrt{8 \pi^3}\sigma_x \sigma_y \sigma_z} \exp\left(-
\frac{p_x^2}{2\sigma_x^2}-\frac{p_y^2}{2\sigma_y^2}-\frac{p_z^2}{2\sigma_z^2}\right)\Big\rangle
\label{ndp}\end{equation} 
where $\sigma_z$ is along the direction of the H bond, and $\sigma_x$ and $\sigma_y$ are in the plane perpendicular to the direction of the H bond, {\it i.e.}. The set of parameters, $\sigma_{x,y,z}$, determine the anisotropy in the momentum distribution lineshape. 
We recall that although the M1 model represents the most general momentum distribution lineshape, it does not allow to directly separate the effects of anharmonicity from those of anisotropy\cite{2011Flammini,2005andreani_adv,2004Reiter_bjp}. 

Top part of Figure \ref{cy_data} shows the $\bar F(y,q)$ function for SW at T=271 K, and its best fit, resulting from model M2. The quality of present data is quite good and the accuracy comparable to previous benchmark experiment on polycrystalline ice\cite{2011Flammini}: the difference between the experimental data and the best fit is within two error bars. This is obtained by the difference between data and fits, both normalized to unity area. The deviations from Gaussian momentum distribution found in SW and ice using model M1 (see the non-zero $c_4$ coefficients in Table \ref{fit_parameters}) can be entirely ascribed to the anisotropy of the momentum distribution. 
The $\sigma_\alpha$ values reported in Table \ref{fit_parameters} unveil a main distinctive feature between SW and ice at $T$ = 271 K: the anisotropy of the momentum distribution is slightly more pronounced in SW than in the solid phase. Similar features are also obtained in previous combined INS and DINS investigation in water at T=285 K and ice at T=271 K\cite{2013andreani}. The difference in the two $n(p)$ lineshapes, magnified by a factor ten, is untangled in the bottom panel of Figure \ref{cy_data}. This shows the radial momentum distributions, 4$\pi p^2 n(p)$, for SW and ice obtained using the M2 model. The high momentum components in $n(p)$ are highly sensitive to, and dominated by, the curvature of the effective hydrogen potential. The anisotropic character of the $n(p)$ is due to the anisotropy of the potential that the hydrogens experience along different molecular directions\cite{2006burnham,morrone_2007,2008Pantalei_prl,Burnham_2008}.

Values of the hydrogen total mean kinetic energy, $\ek = 3\frac{\hbar^2\sigma^2}{2m}$, and the directional components along the three  axes, $\ek_\alpha =\frac{\hbar^2\sigma^2_\alpha}{2m}$, are reported in Table \ref{fit_parameters}. Table \ref{table_comp} compares our $\ek$ values for ice, SW and liquid water with values obtained from INS\cite{2013senesi_OH_str_jcp} and theory\cite{2014moreh}. The $\ek$ value for ice at T=271 K derived in this experiment from M1 together with the $\sigma_\alpha$ values are the same, within the experimental uncertainties, with those obtained in previous measurement at the same temperature\cite{2011Flammini}. This is a reference benchmark which validates and strengthens the total value of $\ek=$156 meV, and its directional $\sigma_\alpha$ components obtained for SW. 

\begin{figure}
\centering\includegraphics[width=14cm]{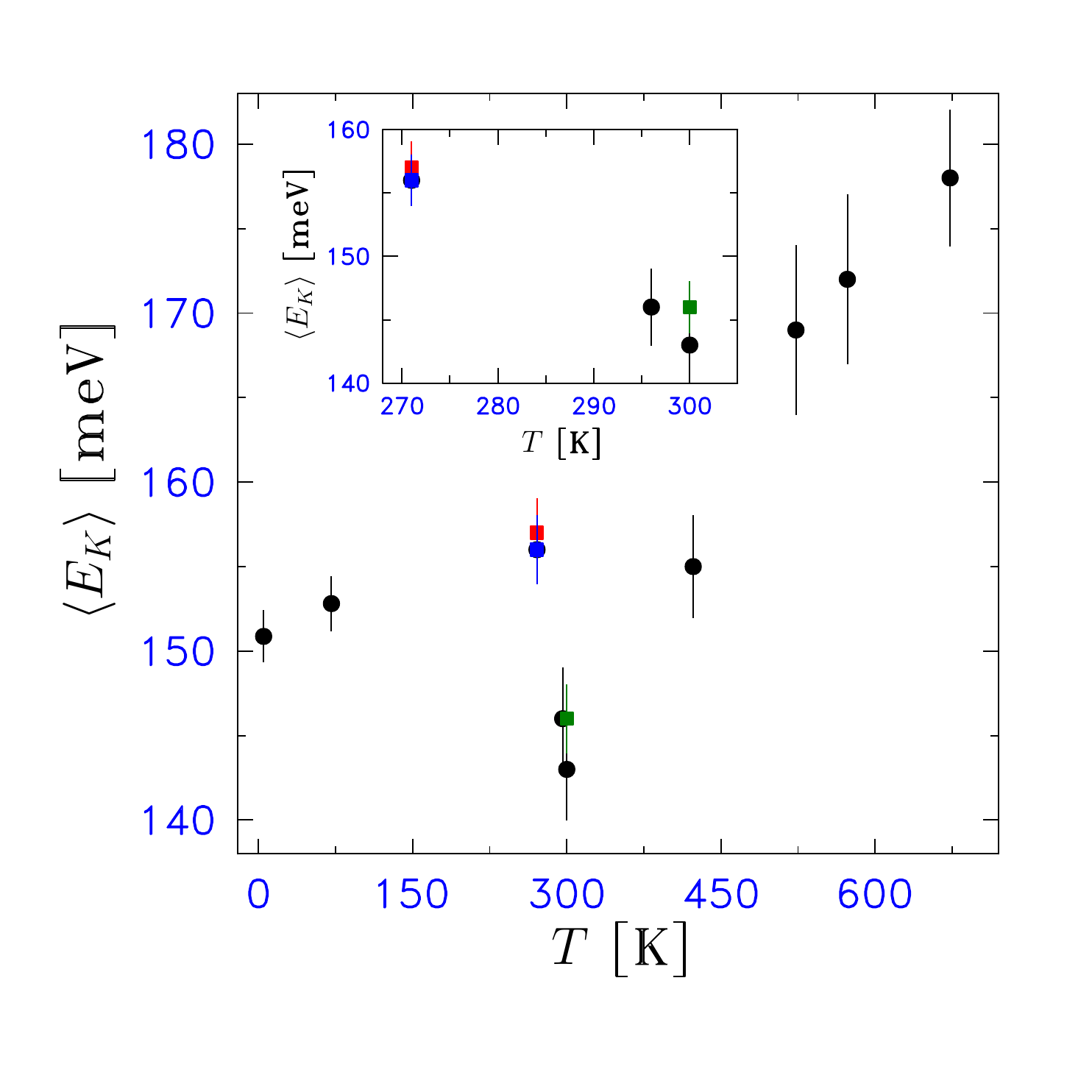}
\caption{(colour online) (colour online) (a) $\ek$ values for SW at T=271 K ({\color{blue}blue full square}), ice at T=271 K ({\color{red}red full square}), and water at 300 K ({\color{green}green full circle}) from present DINS study; (b) $\ek$ values for ice at T=5 K, T= 71 K\cite{senesir_2013} and T=271 K ({\color{black}black full square})\cite{2011Flammini} and water at 300 K and above ({\color{black}black full square}) from previous DINS measurements\cite{2008Pantalei_prl}. The insert shows a magnified picture in the temperature range 270 K - 305 K.}
\label{figure2}
\end{figure}

Several are the experimental investigations showing red-shift in the OH stretching occurring from liquid to solid (see Table \ref{table_comp} and Figure \ref{figure2}), interpreted as a fingerprint of stronger H-bonding in ice respect to water\cite{2011Perakis,andreani_1998}. Changes in the directional components of the $\ek_\alpha$ are used to monitor the competing NQEs associated to phase changes in water near its triple point\cite{2013romanellijpcl}. These reflect the entanglement of the potential energy surface with \ndp that is generated by the uncertainty relation between position and momentum of the hydrogen atom: this is the result of a competition between anharmonic quantum fluctuations of intermolecular bond bending and intramolecular covalent bond stretching. The latter fluctuations strengthens H-bonds whereas the former weakens H-bonds\cite{2013andreani}. Thus in the case of SW and ice the competition between the directional energy components $\ek_\alpha$ is such to produce a subtle cancellation effect, resulting in $\ek$ values close to each other. The DINS technique effectively fingerprints the competition between intra and inter molecular NQEs during the transition from disordered phase (SW) to polycrystalline one (ice). Further result of the present investigation is that $\ek$ value in SW is slightly lower than in ice at the same temperature, in quantitative agreement with Ref.\cite{ceriotti1} and qualitative agreement with Ref.\cite{ramirez2011kinetic}.

As far as the comparison between metastable liquid (SW) and water at room temperature is concerned we find the following results: $\ek$ in SW is about 7$\%$ and 9$\%$ in excess respectively to $\ek$ values in the stable liquid at T=300 K, which are $\ek$= 146 meV from the present study, and $\ek$= 143 meV from Ref.\cite{2008Pantalei_prl} (see Table \ref{table_comp} and Figure \ref{figure2}).  

From Table \ref{table_comp} the experimental result for SW and ice at $T$ = 271 K are in quite good agreement with INS data\cite{2013senesi_OH_str_jcp}; in the case of ice a good comparison is also seen with predictions from an harmonic theoretical model\cite{2014moreh}. Although the harmonic model is successful in reproducing $\ek$ of ice at $T$ = 271 K it overestimates the value of $\ek$ at ambient temperature. This occurs because harmonic models cannot fully account for the softening of the ground state potential energy surface at room temperature in water, in agreement with previous suggestion from a joint DINS and path integral investigation\cite{2008Pantalei_prl}. Results from a path integral molecular dynamics simulation\cite{ramirez2011kinetic} on SW and ice at T=270 K are instead both approximately 8 meV lower than the present experimental results.

\begin{table}
\centering
\begin{tabular}{ccccc}
\hline\hline
Phase &	T [K]	 &	$\ek^{DINS}$ &	$\ek^{INS}$& $\ek^{M}$\\
&&[meV]&[meV] &[meV] \\
\hline
SW		&269	 &	(199$\pm$2)\cite{2008_pietropaolo_prl}	&	(152$\pm$4)\cite{2013senesi_OH_str_jcp}\\
		&270		&	&&(148.1$\pm$0.5)\cite{ramirez2011kinetic}\\
		&271	 &	{\color{blue}156.0$\pm$2.0}		&	(152$\pm$4)\cite{2013senesi_OH_str_jcp}			\\
		&	&	(228$\pm$2)\cite{2008_pietropaolo_prl}&\\
		&273	 &(150$\pm$2)\cite{BZ_2009}&(153$\pm$4)\cite{2013senesi_OH_str_jcp}\\
\hline
Ice		&270 &					&		&(149.5$\pm$0.5)\cite{ramirez2011kinetic}		\\
		&271	 &	{\color{red}157.0$\pm$2.0}	&(158$\pm$4)\cite{2013senesi_OH_str_jcp}&(155$\pm$3)\cite{2014moreh}\\
		&	&  (156$\pm$2)\cite{2011Flammini}	&\\
		\hline	
Liquid		&296		&(146$\pm$3)\cite{BZ_2009}	&(150$\pm$4)\cite{2013senesi_OH_str_jcp}	\\
	&300	 &{\color{green}(146$\pm$3)}	&&(155$\pm$3)\cite{2014moreh}\\
	&&(143$\pm$3)\cite{2008Pantalei_prl}		&		&\\
			
\hline\hline
\end{tabular}

\caption{(colour online) Values of $\ek$ for SW, ice, and water. a) $\ek^{DINS}$ values for SW from present study at T=271 K ({\color{blue}blue}), from path integral molecular dynamics simulation at T=270 K\cite{ramirez2011kinetic} and from previous DINS experiments at T=269 K\cite{2008_pietropaolo_prl}, T=271 K\cite{2008_pietropaolo_prl} and T= 273 K \cite{BZ_2009}; $\ek^{INS}$ values from previous INS experiment at the same corresponding temperatures\cite{2013senesi_OH_str_jcp}; b) $\ek^{DINS}$ for ice from present study at T=271 K ({\color{red}red}), from path integral molecular dynamics simulation at T=270 K\cite{ramirez2011kinetic} and from previous DINS experiments at T=271 K\cite{2011Flammini};  $\ek^{INS}$ value from previous INS experiment at T=271 K and $\ek^{M}$ value from an harmonic theoretical model at T=271 K\cite{2014moreh}; c) $\ek^{DINS}$ for water from present study at T=300 K({\color{green}green}) and from previous DINS experiments at T=300 K\cite{2008Pantalei_prl}, $\ek^{INS}$ and $\ek^{M}$ values for water from previous studies\cite{2013senesi_OH_str_jcp,2008Pantalei_prl,BZ_2009,2014moreh}.}

\label{table_comp}
\end{table}

In this study we derive new quantitative and accurate values for the hydrogen $n(p)$, $\ek$ and directional $\ek_\alpha$ observables in water near the triple point. To the best of our knowledge the present value of $\ek=$156 meV at 271 K in SW represents the most reliable experimental value obtained so far, since it is established through a benchmark measurement on ice at T=271 K, which is correlated with previous DINS result on ice at the same temperature\cite{2011Flammini}. The $\ek$ value in SW is found approximately 10 meV higher with respect to  $\ek$ value for water at room temperature, corresponding to less than 7$\%$ increase. Results from SW and ice at $T$ = 271 K both show \ndp functions with an anisotropic Gaussian lineshapes and directional anisotropic components of the $\ek_\alpha$ tensor. The $\ek$ value in SW can be regarded as an upper limit, which can help to identify those theories that best describe and explain the behaviour of water and hydrogen bonded systems. 

The present DINS study provides a reference set of values of increased accuracy with respect to earlier measurements, yielding new and key insights in the three dimensional potential energy surface experienced by hydrogen in water near the triple point. DINS confirms to be a a unique and sophisticated technique for the investigation the NQEs in water. It is highly desirable to continue to expand the DINS accuracy in order to further refine the cross comparison with the parallel increasing accuracy of path integral molecular dynamics.
 
\section*{Acknowledgement}
This work was supported within the CNR-STFC Agreement (2014-2020) concerning collaboration in scientific research at the ISIS pulsed neutron and muon source. 

\bibliographystyle{unsrt}

\newpage

\end{document}